\documentclass[superscriptaddress,preprintnumbers,nofootinbib]{revtex4}[12pt]
\usepackage{graphicx}
\usepackage {amsmath, amssymb, graphicx, multirow, rotating}
\usepackage{hyperref}

\begin{document}

\def\simgt{\mathrel{\lower2.5pt\vbox{\lineskip=0pt\baselineskip=0pt
           \hbox{$>$}\hbox{$\sim$}}}}
\def\simlt{\mathrel{\lower2.5pt\vbox{\lineskip=0pt\baselineskip=0pt
           \hbox{$<$}\hbox{$\sim$}}}}
\newcommand{\met}{\mbox{${\rm \not\! E}_{\rm T}$}}

\newcommand{\Ldet}{l__{\rm detect}}
\newcommand{\Ldist}{R}
\newcommand{\Csun}{C_\odot}
\newcommand{\Cearth}{C_\oplus}
\newcommand{\unit}[1]{\,\textrm{#1}}
\newcommand{\sigSI}{\sigma_{\rm SI}}
\newcommand{\sigSD}{\sigma_{\rm SD}}
\newcommand{\mdm}{m_{\rm DM}}
\newcommand{\mgauge}{m_{\gamma_d}}
\newcommand{\param}[3]{\left(\frac{#1}{#2}\right)^{#3}}
\newcommand{\mlolip}{m_{\rm LOLIP}}

\preprint{YITP-SB-09-32}

\title{Searches for Long Lived Neutral Particles}

\author{Patrick Meade}
\affiliation{C.N. Yang Institute for Theoretical Physics, Stony Brook University, Stony Brook,  NY 11794}
\affiliation{ School of Natural Sciences, Institute for Advanced Study,
 Princeton, NJ 08540}
 
\author{Shmuel Nussinov}
\affiliation{Tel Aviv University, Sackler School Faculty of Sciences, Tel Aviv 69978, Israel and Schmid Science Center Chapman University, Orange California 92866}

\author{Michele Papucci}
\affiliation{ School of Natural Sciences, Institute for Advanced Study,
 Princeton, NJ 08540}
 
 \author{Tomer Volansky}
\affiliation{ School of Natural Sciences, Institute for Advanced Study,
 Princeton, NJ 08540}

\begin{abstract}
  An intriguing possibility for TeV scale physics is the existence of
  neutral long lived particles (LOLIPs) that subsequently decay into SM states. Such particles are many cases indistinguishable from missing transverse energy (MET) at colliders. 
  We propose new methods to search for these particles using neutrino telescopes. We study their detection prospects, assuming production either at the LHC or through dark matter (DM) annihilations in the Sun and the Earth.
We find that the sensitivity for LOLIPs produced at the LHC is limited by luminosity and detection energy thresholds. On the other hand, in the case of DM annihilation into LOLIPs, the sensitivity of neutrino telescopes is promising and may extend beyond the reach of upcoming direct detection experiments. In the context of low scale hidden sectors weakly coupled to the SM, such indirect searches allow to probe couplings as small as $10^{-15}$.
\end{abstract}

\maketitle

\section{Introduction}\label{sec:introduction}
Current and future experiments are beginning to probe the TeV scale. This energy scale incorporates the mechanism for Electroweak symmetry
breaking (EWSB) and quite possibly the solution to the dark matter (DM) mystery. Ê While many theoretical possibilities for TeV scale physics have been explored in great detail, it is still quite possible that something truly unexpected will be discovered. Ê ÊThe reason for this is because no experimental evidence thus far that favors a particular model proposed so far. The possibility for exotic experimental signatures from beyond the SM physics has been recently explored in more detail in an attempt to ensure that nothing is missed at the LHC. ÊHidden Valleys~\cite{Strassler:2006im}, Quirks~\cite{Kang:2008ea} and other theories have demonstrated the limitations of current experimental searches at colliders by exploring the possibility that new physics might not be triggered on, or easily reconstructed at the LHC.

One intriguing possibility for new physics are particles which have very long lifetimes. ÊIf these particles are charged under the SM, they are typically known as charged massive particles (CHAMPs).  These have striking experimental signatures, but are straightforward to search for. ÊIf the LOLIPs are neutral, experimental searches are much more difficult and strongly depend on their lifetime. ÊIn this paper we will explore new ways to search for neutral LOLIP decays that allow one to explore lifetimes that were previously beyond the experimental reach.

Neutral LOLIPs are quite generic in models of physics beyond the SM. ÊFor instance, additional gauge sectors or new particles that couple only very weakly to the SM can easily have macroscopic lifetimes, $\tau \gg \mathcal{O}(10^{-6})\,\mathrm{s}$. ÊThe possibility for such lifetimes has already been discussed in many specific contexts before. ÊFor instance, in theories of Êgauge mediated supersymmetry breaking (GMSB), the decay length of the next to lightest super-partner (NLSP) into the gravitino is macroscopic when the primordial supersymmetry (SUSY) breaking scale is $\sqrt{F}\gtrsim 1000\,\mathrm{TeV}$. ÊAnother scenario, that of a low scale hidden sector, has received considerable attention during the last year. Ê Indeed, indirect clues from astrophysics have motivated studies of these ``unconventional" models at the TeV scale. ÊThe recent PAMELA~\cite{pamela} and Fermi~\cite{fermi} measurements imply that if the experimental signals are a consequence of DM annihilations, DM must have non-standard properties~\cite{us}, suggesting the possible existence of an additional sector, weakly coupled to the SM~\cite{nima}. ÊIf this is the case, the low lying states in the hidden sector may be long lived and decay to the SM through some small coupling, $\epsilon$. 

Recently several ways have already been proposed to directly search for neutral LOLIP decays when the decay length is not much longer than the size of the detector used. Ê These fall under the general categories of using high energy colliders~\cite{ArkaniHamed:2008qp, Han:2007ae, feng}, low energy $e^+e^-$ colliders~\cite{Essig:2009nc,reece,Aubert:2009pw}, fixed target experiments~\cite{Bjorken:2009mm,reece, pospelovFT}, and rare meson decays~\cite{reece}. ÊWhile many new ideas for searching for neutral LOLIPs have been put forth, the lifetimes probed are relatively short and correspondingly the coupling $\epsilon$ is typically not smaller than $10^{-7}$ at most.

An even more intriguing and less studied possibility is the case when the decay length of a neutral LOLIP is much greater than
the size of the detector used to produce and study it. ÊFor instance if neutral LOLIPs are produced at the LHC and have an effective decay length significantly larger than the size of ATLAS or CMS, they will simply escape the detector and be recorded as MET. ÊFrom this perspective one could not distinguish these particles from stable particles.  Experimental ways to differentiate between these two possibilities are very limited. ÊTwo indirect methods have been suggested for probing a hidden sector that has particles with very long lifetimes: (i) looking at the effects on the evolution in stars, for instance supernova cooling~\cite{Bjorken:2009mm} and (ii) measurements of certain gamma ray features in models related to DM~\cite{Ruderman:2009ta,Ruderman:2009tj}. ÊThese methods are very model dependent and in this paper we will present more generic methods that allow one to search for LOLIPs with decay lengths much greater than $\mathcal{O}(\mathrm{km})$ .

There are three ingredients which dictate the possible reach of a search for LOLIPs. ÊThe first is the production mechanism and its rate. ÊThis mechanism will also determine the momentum distribution of the LOLIPs, which may influence the capability to detect this particle. ÊThe second ingredient is the lifetime of the particle, which determines the probability of the LOLIPs to decay inside the detector. Ê ÊThe physical quantity of interest is the effective lifetime of the LOLIPs, $\tau_{eff}=\gamma \tau$, which depends on the typical momenta of the produced particle. Ê ÊFinally, the final states into which the LOLIP decays, will determine the types of detectors that are most applicable and how these particles can be searched for.

In this paper we will focus on two mechanisms for production. ÊThe first, and most straightforward perhaps, is to
produce the particle at a collider. ÊIf the LOLIP weakly couples to the SM but has a low mass, then in principle high
luminosity low energy experiments can produce it~\cite{reece}. ÊAlternatively, the LOLIP could have a
larger mass, as is the case for instance, of a bino NLSP in gauge mediation. Perhaps more interestingly, the long lived state might be in another sector that is only effectively accessed by producing heavier states such as a Z' or supersymmetric particles, which subsequently decay into this new sector. ÊIn this case luminosity is still important, but higher energy is critical for producing these new states. ÊThus the most effective way for producing Êthese particles is with the LHC. Ê

The second method, which turns out to be more promising, is producing LOLIPs from DM annihilation. Ê For this production mechanism to be useful, one has to focus on regions where the DM density is high, such that a large enough flux is generated. ÊThe most interesting regions that are potentially useful are the Sun and Earth. ÊÊGiven the association with DM, there are a number of already existing constraints on this scenario, coming from Fermi~\cite{fermi}, Milagro~\cite{Atkins:2004qr}, and SuperKamiokande~\cite{upgoingmuons} that Êexclude part of the parameter space.

For either production mechanism, we show that using large volume neutrino telescopes is the most powerful method for increaseing the reach to longer lifetimes and weaker couplings . ÊAs we will demonstrate, these experiments can probe hidden sectors with $\epsilon$ as small as $10^{-15}$, and lifetimes comparable to an Astronomical Unit, in the case of production through DM. ÊWe also demonstrate that detecting the decay of LOLIPs produced at the LHC is very difficult even with the improved upcoming experiments. ÊHowever we suggest new methods that may be improved upon, and in addition put forth the idea of new detectors dedicated to these searches that will be investigated further in~\cite{workinprogress}. 

The rest of the paper is structured as follows. ÊIn Section~\ref{sec:models} we begin by discussing specific examples of models that contain neutral LOLIPs. ÊWe give the parametric dependence of their decay lengths, which is necessary to translate into specific model parameters the reaches that we present as a function of lifetime. ÊIn Section~\ref{sec:experiments} we give the details for the neutrino telescope experiments that we will use to bound neutral LOLIP decays and to search for them. ÊWe include the details of how neutral LOLIP decays differ from the standard events looked at by these experiments and quantify this difference for certain cases. Ê In Section~\ref{sec:lhc} we discuss potential methods for discovering neutral LOLIPs produced at the LHC. ÊIn Section~\ref{sec:sunandearth} we discuss the bounds and the discovery potential for neutral LOLIP decays coming from DM annihilation in the Earth and Sun. ÊIn that Section, we also discuss the bounds from the Fermi and Milagro experiments in the case that LOLIPs are produced from DM. ÊFinally in Section~\ref{sec:conclusions} we discuss future avenues for LOLIP searches and the ramifications of the bounds and reaches Êwe calculated for individual models.

\section{Models}\label{sec:models}
Given that the experimental results discussed in Sections~\ref{sec:lhc} and ~\ref{sec:sunandearth} depend only on the decay length and the production rate, it is useful to have several benchmarks models that can accommodate the long lifetimes explored in this paper.
In this Section we present some examples, providing the relevant formulae to relate these models to our general results of Sections~\ref{sec:results}. We do not attempt to give a complete list of all models that contain LOLIPs, but the following are rather generic examples that can easily have decay lengths in the 1 km to $10^{15}$ km range.

 \begin{enumerate}
 \item[I.]{\bf Gauge Mediation.} Ê In this case the NLSP can decay to the
 Êlight Êgravitino, $\tilde G$, which is typically assumed to be the DM particle. Ê The lifetime is determined
parametrically by the SUSY breaking scale $F$ and the mass of the
NLSP, $m_{\rm NLSP}$,
 \begin{eqnarray}
\label{eq:5}
c\tau_{\rm NLSP} \simeq
\left(\frac{m_{\rm NLSP}^5}{16 \pi F^2} \right)^{-1}
= 10 \unit{km} 
\left(\frac{100\unit{GeV}}{m_{\rm NLSP}} \right)^5
\left(\frac{\sqrt{F}}{10^4\unit{TeV}} \right)^4\, .
\end{eqnarray}
The lifetime can span many orders of magnitude, depending on the value
of $F$. ÊUnless the theory is complemented with a different DM candidate which is allowed to annihilate into the NLSP, long lifetimes in GMSB models may only be discovered using the LHC. The case where such extra DM candidate is introduced does not present qualitative differences to the case (III-d) discussed in the following and we will not discuss it further here.
Finally, if one relaxes the requirement of a neutral LOLIP, another intriguing possibility present in GMSB is to have charged LOLIPs such as stau NLSPs. ÊIn this case very long lived staus with decay lengths of several km can also be detected in neutrino telescopes from production via high energy cosmic ray neutrinos~\cite{staunlsp}. 

\item[II.]{\bf R-parity Violation.} ÊR-Parity Violation in SUSY models is another typical scenario where very-long lifetimes may be obtained~\cite{rpv}. However, as in the case of GMSB models, the R-parity violating MSSM does not contains long lived particles in the DM annihilation final states. It becomes therefore relevant for the study performed in this paper only if another DM candidate is introduced in the theory. Moreover, the parametrics of the LSP lifetime strongly depends on the particular operator used to break R-Parity and many possibilities are present. Ê

\item[III.] {\bf Hidden sector with a Vector Portal.} ÊAs discussed in the introduction, this theory has been extensively studied in the context of the cosmic ray anomalies. ÊThe hidden sector is assumed to communicate with the visible sector through gauge kinetic mixing with the SM hypercharge,
\begin{eqnarray}\label{eq:6}
\epsilon_V F_{\mu\nu}^\prime B^{\mu\nu}\,.
\end{eqnarray}
Here $F^\prime_{\mu\nu}$ is the field strength of the gauge field(s) in the hidden sector and we have suppressed the adjoint index in the case of a non-Abelian group. Ê For Abelian hidden sectors, $\epsilon_V$ is expected to be of order $10^{-2}-10^{-4}$ since the operator above is renormalizable and may possibly arise at the one loop level.  On the other hand, for non-Abelian gauge groups, $\epsilon_V$ may be significantly smaller. Indeed, in that case, Eq.~\ref{eq:6} arises from a non-renormalizable operator such as,
\begin{equation}
\frac{\Phi^a}{M}F_{\mu\nu}^a B^{\mu\nu}\,,
\end{equation}
where $M$ is some high scale and $\Phi^a$ is in the adjoint representation of the hidden gauge group.  Depending on the ratio $<\Phi^a>/M$,  $\epsilon_V$ can be as small as $10^{-16}$ (corresponding to $M$ at the GUT scale and the VEV of $\Phi$ at the GeV scale).  In other examples, even higher dimensional operators may produce miniscule couplings.    We conclude that $\epsilon_V$ can span a wide range of values.

The lowest lying state in the hidden sector may be either stable or long lived Êwith a lifetime enhanced with negative powers of $\epsilon_V$. ÊWe consider four possibilities:

\begin{enumerate}
\item {\it Hidden photon,, $\gamma_d$. $\gamma_d\rightarrow f \bar f$}: ÊIf decays into the hidden sector are kinematically forbidden, the gauge field, $\gamma_d$, will predominantly decay into a pair of SM fermions through the kinetic mixing. The branching fraction of the various channels depends on the mass of gauge fields, $\mgauge$. ÊThe decay length is found to be~\cite{Batell:2009yf},
\begin{eqnarray}\label{eq:7}
c\tau_{\gamma_d\rightarrow \bar ff} = \left(\frac{1}{3}\alpha_{\rm EM} \epsilon_V^2 \mgauge\left(1+\frac{m_f^2}{m_{\gamma_d}^2}\right)\beta_f\right)^{-1} \simeq 0.8 \unit{km} Ê\param{10^{-8}}{\epsilon_V}{2} \param{1\unit{GeV}}{\mgauge}{}\,, Ê
\end{eqnarray}
where $\beta_f = (1-4m_f^2/m_{\gamma_d}^2)^{1/2}$ is the phase space factor which we take to be 1 for the numerical estimate. We see that lifetimes in the range $1 \unit{km}< c\tau < 10^{15}\unit{km}$ considered here require values for $\epsilon$ of order $10^{-8}\div10^{-15}$. Ê 
 
\item {\it Hidden Higgs, $h_d$. $h_d\rightarrow f\bar f$}: ÊIf the hidden higgs is the lightest state in the hidden sector, it predominantly decays to SM fermions at 1-loop, with the approximate Êdecay length~\cite{Batell:2009yf}
\begin{eqnarray}\label{eq:8}
c\tau_{h_d \rightarrow \bar ff} \simeq\left(\frac{\alpha_d\alpha_{\rm EM}\epsilon_V^4}{2\pi^2}\frac{m_f^2}{\mgauge^2}m_{h_d}\beta_f^3\right)^{-1}\simeq 0.7 \times 10^7 \unit{km} \param{10^{-4}}{\epsilon_V}{4}\param{0.01}{\alpha_d}{}\param{\mgauge}{1 \unit{GeV}}{2} \param{1\unit{GeV}}{m_{h_d}}{}\,.
\end{eqnarray}
$\alpha_d=g_d^2/4\pi$ is the gauge coupling in the hidden sector and we neglected an $O(1)$ factor coming from the loop integral. Here $m_f$ is the fermion mass, taken for concreteness to be the muon mass. Ê As opposed to the previous case, the lifetimes of interests are obtained for relatively large values of $\epsilon_V\sim 10^{-5}\div10^{-3}$. ÊWe note, that if the higgs is not the lightest particle, but has a mass $\mgauge < m_{h_d} < 2\mgauge$, it can decay through one off-shell hidden photon, in which case its lifetime is proportional to $\epsilon_V^{-2}$ but is still larger than the one of the hidden photon due to the 3-body phase space suppression.

\item {\it Hidden Gaugino, $\tilde \gamma_d$. $\tilde \gamma_d \rightarrow \tilde G \gamma$} ÊIn the supersymmetric scenario, the hidden gaugino may be degenerate with or lighter than the hidden gauge boson. ÊIf in addition the gravitino is light, as in the case of GMSB, the only possible decay mode for the gaugino is into a gravitino and a SM photon, $\tilde \gamma_d \rightarrow \gamma + \tilde G$~\cite{Ruderman:2009ta,Ruderman:2009tj}. The lifetime for this decay is
\begin{eqnarray}\label{eq:9}
c\tau_{\rm \tilde \gamma_d\rightarrow \gamma\tilde G} \simeq \epsilon_V^{-2}\left(\frac{m_{\tilde \gamma_d}^5}{16 \pi F^2} \right)^{-1} \simeq 3\cdot 10^6 \unit{km} \param{10^{-4}}{\epsilon_V}{2} \left(\frac{50\unit{GeV}}{m_{\tilde\gamma_d}} \right)^5 \left(\frac{\sqrt{F}}{1000\unit{TeV}} \right)^4\, .\end{eqnarray}
Note the large gaugino mass required to produce the lifetimes of interest. ÊWhile the small mass limit of this model may be constrained indirectly by experiments such as Fermi~\cite{Ruderman:2009ta}, large masses in the hidden sector are hard to probe and have not been constrained so far. ÊThe ability to probe such models demonstrate the strength of the indirect measurements proposed here. ÊFinally we comment that this scenario can be generalized to allow other light fermions in the spectrum.

\item {\it Hidden Gaugino, $\tilde \gamma_d$. $\tilde \gamma_d \rightarrow \tilde G \gamma_d$}: 
Another possibility is that the hidden spectrum allows the hidden gaugino to decay into a real hidden gauge boson. The parametrics of this decay is the same as in case (I) and the hidden gaugino may be very long lived. Since the long lifetime of $\tilde\gamma_d$ is entirely accounted for by $\sqrt F$, $\epsilon_V$ can be quite large and the subsequent $\gamma_d$ decays can be prompt, resulting in $\tilde\gamma_d$ producing final states containing SM fermions. This case presents a detection scenario very similar to (III-a), but it is achieved without the need of minuscule $\epsilon_V$. 
\end{enumerate}

 \item[IV.] {\bf Hidden sector with a Higgs portal.} Ê Another possibility for the hidden sector to communicate with the SM is through the so called Higgs portal, with an interaction of the form
\begin{eqnarray}\label{eq:10}
\lambda H^\dagger H h^{\prime\dagger} h^{\prime}\,.
\end{eqnarray}
$h'$ can then decay to SM fermions through the Higgs couplings. ÊDefining the effective coupling, 
\begin{equation}
\epsilon_H = \lambda \frac{\langle h^\prime\rangle v}{ m_h^2},
\end{equation}
with $v=174\unit{GeV}$ the vev of the SM Higgs and $m_h$ its mass, the lifetime is found to be,
\begin{eqnarray}\label{eq:11}
c\tau_S \simeq \epsilon_H^{-2} \left(\frac{G_F m_f^2}{4\sqrt{2}\pi}\beta_f^3 m_{h^\prime}\right)^{-1} \simeq 30\unit{km}\param{10^{-6}}{\epsilon_H}{2}\param{1\unit{GeV}}{m_{h^\prime}}{}\,,
\end{eqnarray}
where we have assumed again a decay into muons. ÊThe size of $\epsilon_H$ here is required to be small to allow for a sizeable lifetime. ÊThis is obtained in supersymmetric models where the interaction (\ref{eq:10}) may be generated Êin the K\''ahler potential at a high scale $\Lambda$. ÊIn such a case, $\lambda$ is expected to be of order $\lambda \sim m_{h^\prime}^2 / \Lambda^2$ which can naturally be small.
\end{enumerate}

In order to detect the above particles coming from the Sun or the Earth, the DM needs to interact with and annihilate into the LOLIPs. In the case of hidden sector models, if the DM is not directly coupled to the visible sector, it interacts with SM matter and, as a consequence with nucleons, only through the corresponding portal. ÊSuch interactions are crucial to allow for the capture of DM in the Sun and Earth. ÊIn the vector portal case, the DM-nucleon cross-section is given by~\cite{nima},
\begin{eqnarray}
 Ê\label{eq:12}
 Ê\sigma_{\chi n} = \frac{16\pi Z^2 \alpha_{\rm EM}\alpha_d \epsilon_V^2
 Ê Ê\mu_{\chi n}^2}{A^2 \mgauge^4} \simeq 3\times
 Ê10^{-39}\unit{cm}^2 \param{\epsilon_V}{10^{-4}}{2}\param{1\unit{GeV}}{\mgauge}{4}\,,
\end{eqnarray}
where $\mu_{\chi n}$ is the reduced DM-nucleon mass and we have taken $A=2Z$. 

Below, for models (III-b), (III-c) and (III-d), we will assume the above cross-section. ÊWhile these models are excluded for $\epsilon \gtrsim 10^{-6}$ in the case of elastic scattering of DM off nucleons~\cite{cdms, xenon10}, they are not excluded in the Êinelastic case with mass splittings of order $\sim 150\unit{keV}$ or above. ÊOn the other hand, in models (III-a) and (IV), the relevant $\epsilon$ is typically too Êsmall to allow for both long lifetimes and sufficiently large scattering cross-section (and therefore capture rates). Ê It is possible, however, that the DM is coupled both to the hidden sector and visible sector with sizable annihilation fractions to both. ÊThis would be the case, for example, if the DM transforms under the gauge groups of both sectors. ÊAs a consequence, the ÊDM-nucleon cross-section may be dominated by ÊZ-exchange and independent of $\epsilon$. ÊBelow, we will assumeÊthis possibility for models (III-a) and (IV).

\section{Experiments and Signatures}\label{sec:experiments}
 
In this section we describe some of the experimental details needed to extract the bounds on LOLIP decays as well as their potential reach. ÊAs mentioned in the introduction, we focus on neutrino telescopes, and in particular we consider SuperKamiokande~\cite{upgoingmuons}, IceCube~\cite{icecube}, ANTARES~\cite{AntaresTDR}, and the proposed km$^3$ neutrino telescope in the Mediterranean KM3NeT~\cite{KM3NeTCDR}. ÊSince we will discuss measurements from the Sun, we also consider bounds from the Fermi~\cite{fermi} and Milagro~\cite{Atkins:2004qr} experiments, which we describe in more detail in Section~\ref{sec:existing-constraints}.

The probability for a particle with a decay length $L = \gamma c \tau$, to decay inside a detector of size\footnote{The quantity $d$ is a combination of the actual size of the detector and of the typical range traveled by the particles being detected, as in the case of muons.} $d$ is given by,
\begin{eqnarray}\label{surviveanddecay}
 Ê\label{eq:1}
 ÊP_{\mathrm{decay}}=e^{-D/L} \left(1-e^{-d/L}\right) \simeq \left\{\begin{array}{ccl} 
 Êe^{-D/L} && L \ll D 
 Ê\\
 Ê\frac{d}{L} && L \gtrsim D \end{array}\right. \,.
\end{eqnarray}
Here $D\gg d$ is the distance between the detector and the source. In the second equality above, we considered the two important limits, demonstrating that if the decay length is comparable to or larger than the distance to the source, the probability to observe the particle increases linearly with the size of the detector. ÊConversely, if the LOLIP is short lived (compared to $D$), the probability for decaying inside the detector is exponentially small. Ê
$P_{\rm decay}$ is maximized when $D =L$. Ê
As a consequence, LOLIPs can only be detected if produced at relatively nearby sources, such as the LHC, the Earth or the Sun. ÊLooking for LOLIPs from DM annihilation in the Galactic center (GC) faces a suppression of at least Ê$P_{\rm decay}^{\rm GC} \sim 10^{-17} d/\mathrm{km}$, which renders this source unusable for detection or for setting limits. Ê
Besides $P_{\mathrm{decay}}$ the other dependence on $d,D$ comes in from the usual solid angle suppression $\Delta\Omega=A_{eff}/D^2$, so that when $P_{\rm decay}$, Êis maximized the total suppression is approximately $V_{\det}/L^3$ where $V_{\det}$ is the detector volume.

We focus on neutrino telescopes, since these are the largest Êvolume detectors currently available. 
These telescopes are typical arrays of photomultiplier tubes (PMTs) in various configurations. ÊWhile many different types of particles provide signatures in these telescopes, yet given the large volumes and coarse structure the events can be roughly subdivided into two categories. Muons, passing through a large part of the detector, leave a ``track"-like signature, mostly in the form of Cherenkov radiation. ÊOn the other hand, any hadronic or electromagnetic particles (e.g. quarks, electrons, photons), provide more localized (typical size of $O(10\,\mathrm{m})$) sources of light since they travel a much shorter distance than muons. ÊWith this basic categorization in mind, we focus on Ê$e^\pm/\gamma$ or $\mu^\pm$ final states as typical representatives of these two categories (we do not consider $\tau$s Êwhich may produce the intermediate situation of a ``spot+track'').

Before describing each of the experiments, a brief discussion on the case of $\mu^{\pm}$ final states is in order and will apply to all the Cherenkov detectors considered below. Muon pairs originating from DM annihilations into LOLIPs are typically highly boosted. ÊGiven the boost factors involved, these muons will not be separated enough to be identified as two individual muons: for an $O(100\,{\rm GeV})$ muon pair produced by a LOLIP with a mass of $1\,{\rm GeV}$, the typical separation between the tracks is a few meters at most, which translates into O(10ns) relative time delay for the emitted Cherenkov light. Such time is comparable to typical readout times of PMTs. ÊUnless specific analyses are performed, these di-muons will most of the time be recognized as a single muon. However the Cherenkov light yield along the track of a di-muon is higher (factor of 2) than a single sub-TeV muon. Since the Cherenkov light yield is almost independent of the energy of the particle, such di-muons will be recognized as muons above their critical energy (where they start loosing energy by radiative processes). The critical energy in water or ice is of O(700GeV). This fact provides a good handle for reducing the backgrounds since the atmospheric neutrino spectra are rapidly falling with the energy. Moreover, by studying at the differential light yield along the track, a di-muon event will look different from a single muon above the critical energy: for a very energetic muon the additional light emitted on top of the Cherenkov emission comes from radiative processes which are most effective at the beginning of the track and quickly die off. On the other, hand a di-muon will yield the additional light continuously along the track, providing another possible handle for isolating the signal. ÊIt would be very interesting to study this effect further and the implications it may have for energy measurements in neutrino telescopes~\cite{superkenergy}. 

Turning to the specific experiments, SuperKamiokande is a large volume of water surrounded by PMTs. ÊWe can use its data from upgoing muons to constrain LOLIPs from the Earth and the Sun. These LOLIPs may either decay into muons in SuperK, or decay earlier, producing neutrinos via subsequent muon decays. These interact with the material and produce muons that will be detected. ÊIt is important to note that looking at the combined total flux provides only a rough bound because, as  explained above, the events that come from LOLIPs which decay inside the detector are quite different from standard charged current neutrino scattering. 

The ANTARES experiment is a more modern neutrino telescope and is a test project for a future km$^3$ array in the Mediterranean, KM3NeT. ÊWhile this experiment is not the largest neutrino telescope, it is located approximately 300 km from the LHC which makes it the most compelling detector when discussing LOLIP production at the LHC in Section~\ref{sec:lhc}. Ê Ê The experiment is comprised of 12 strings at a depth of approximately $2500\,$m underwater in the Mediterranean sea. Ê Each string has 25 storeys separated vertically by $14.5\,$m and the strings are spread out horizontally from each other by approximately 70 m. ÊEach storey contains an array of three large PMTs. ÊThe effective area for this experiment is $\sim 0.1\,$km$^2$ for very high energies and drops rapidly towards the 100 GeV range where the effective area is of order of $0.02\,$km$^2$. ÊFor a LOLIP that decays into $\mu^+\mu^-$, we have to take into account that the muon pair has a very small opening angle and there therefore an increased light yield which in turn affects the effective area, $A_{eff}$, Êcompared to a single muon event. Ê We have done a Monte Carlo simulation of this effect taking into account the ANTARES geometry and trigger table (as well as other effects like the PMT response as in or the water attenuation of light) and we nominally find that the effective area of ANTARES for a $\mathcal{O}(100\,\mathrm{GeV})$ horizontal di-muon is roughly 3 times larger than the case of a single muon at trigger level.

When discussing the potential reach for neutral LOLIPs, we will also examine the planned KM3NeT telescope. Ê This turns out to be more interesting for potential LHC LOLIP production if KM3NeT is built at the site of ANTARES. ÊKM3NeT is a proposed km$^3$ neutrino telescope to be built at one of several locations in the Mediterranean sea. ÊTo determine its possible effectiveness, we used the proposed ANTARES-like design for it. ÊThis consists of a homogeneous array of 225 strings separated by 95 m from one another, with 37 storeys each spaced 15.5 m apart. ÊFor simplicity we use the original optical module design of ANTARES and the same triggers to MonteCarlo the detector response. Ê ÊWe find that for muon energies of Ê$\sim 100\,$GeV the effective area for KM3NeT is about a factor of 30 larger for LOLIP decays than ANTARES at the same energy. The KM3NeT effective area also sharply falls below $100\,$GeV based on our simulations, the main reason being the large spacing between the strings.

The largest available neutrino detector currently is the IceCube detector. ÊIceCube~\cite{icecube} is a km$^3$ detector which consists of 80 strings at depths between $1400\,$m and $2400\,$m under the Antarctic ice. ÊThe strings are distributed over an area of approximately $1\,$km$^2$ and each contains 60 optical modules spaced 17 m apart. ÊFor this experiment we Êuse the effective areas given in~\cite{GonzalezGarcia:2005xw}. ÊWe also investigate the inclusion of the DeepCore extension (a more densely instrumented inner volume designed to reduce the energy threshold of IceCube), by correcting these effective areas according to the preliminary information of~\cite{0907.2263}. This does not take into account the effects associated with the special type of events that LOLIPs produce with two highly collimated charged particles. ÊThe reach we give in Section~\ref{sec:sunandearth} will thus be conservative. ÊIt would be interesting to determine exactly how sensitive neutrino telescopes are to these types of events.

\section{LHC Production}\label{sec:lhc}

When LOLIPs are produced at the LHC, some fraction of them will decay within the volume of the ATLAS or CMS detectors. ÊGiven a canonical size for these detectors of approximately $d\sim10\,$m, and assuming a decay length $L$, the amount of particles that decay within the detector is given by,
\begin{equation}
N_{decayed}=N_{produced} \left(1-e^{-d/L}\right),
\end{equation}
which, as in Eq.~\eqref{eq:1} gives
\begin{equation}
N_{decayed}\approx N_{produced}\frac{d}{L}.
\end{equation}
This also holds if only part of the detector can be used for detecting the LOLIP decays, as in the more realistic size of $1\,$m. ÊTaking a typical strong production cross-section of order ${\cal O}(100\unit{pb})$ at the LHC for heavy particles, one obtains  $N_{produced}\sim 10^6$ events with an integrated luminosity of $10\,\mathrm{fb}^{-1}$. ÊConsequently, for a decay length of order a few hundred meters,  thousands of LOLIPs may decay inside the LHC detectors. ÊThere has been a great deal of work on the possibility of searching for highly displaced vertices at both ATLAS and CMS. However, given the many backgrounds, it is not clear how many events one would need for the LHC detectors to be useful in studying neutral LOLIP decays. ÊMoreover, for lifetimes comparable or greater than $\mathcal{O} (100\,\mathrm{km})$, the number of events at the LHC becomes negligible and the LOLIPs will only show up as MET.

In order to differentiate between stable particles and LOLIPs external detectors are needed. ÊThe most straightforward idea would be to use other detectors to conduct a long baseline experiment similar to MINOS~\cite{MINOS}. The nearest large volume detector to the LHC is the ANTARES neutrino telescope in the Mediterranean sea. 
A rough estimate of the detection rates at ANTARES is given by the number of particles decaying inside the ANTARES detector. If we assume the effective lifetime of these particles is of $\mathcal{O}(300\,\mathrm{km})$, and assuming they are sufficiently energetic such that $A_{eff}\sim 0.1\,{\rm km}^2$, one has
\begin{equation}
N_{decayed}\sim N_{produced}\frac{d}{c\tau_{eff}}\frac{A_{eff}}{4\pi\, D^2} \sim \frac{0.1}{300}\frac{0.1}{4\pi\, 300^2} \sim 10^{-10} N_{produced}.
\end{equation} 
This means that for cross-sections below $\mathcal{O}(100\,\mathrm{nb})$ ANTARES will not be useful for detecting such LOLIP decays. Ê If the KM3NeT experiment was located at the same site as ANTARES, it would in principle have a better chance to discover LOLIP decays. ÊTo illustrate the utility of KM3NeT we choose a benchmark $Z'$ candidate with a 500 GeV mass that decays into two LOLIPs which subsequently decay into a pair of muons. ÊThe $Z'$ model point we chose has a production cross-section of $12\,$pb. Ê If we choose a lifetime $c\tau=10^5\,$cm to minimize the suppression of the decay probability, we find that for the optimal orientation of the KM3NeT,  $1.5\times10^{-2}$ events will be observed per year.  This should be compared with $6\times 10^{-4}$ events will be observed per year for ANTARES. ÊThe event rate for this benchmark is clearly too low. ÊOther models such as GMSB where one could use the entire SUSY cross-section would be more promising, however with the current LHC designed luminosity, this is doubtful as well since KM3NeT will only be sensitive to $\mathcal{O}(\mathrm{nb})$ event rates.  ÊIt is possible that Super-LHC (for a review see~\cite{Mangano:2009ph}) may be sufficient to allow for a detection of LOLIPs at KM3NeT.  We postpone this question to future work.  

The low number of events can be understood by noting the tension between the need to raise the collider energy in order to increase the effective area and the need to lower the energy in order to increase the production rate.  It would therefore be advantageous to lower the threshold in  KM3NeT as much as possible. ÊDespite the low event rates, it is expected that the background in these neutrino telescopes can be removed entirely with the use of angular information and timing synchronization with the LHC detectors.  Thus observing any $\mathcal{O}(1)$ number of events could be a clear signal of LOLIP decay.

Given the small rates associated with the nearest neutrino telescopes one is forced to examine other alternatives for studying LOLIP decays produced at the LHC. ÊThe suppression due to the solid angle clearly indicates that new detectors that are closer to the LHC are needed.  Additionally one could attempt to use the existing  
 LHC detectors as long baseline detectors for each other. ÊEvents at ATLAS could be recorded at CMS or vice versa. ÊThese possibilities require detector design studies, and a more detailed analysis of timing between the experiments and the triggers. ÊWe leave both possibilities for future work~\cite{workinprogress}.

\section{LOLIPs from the Sun and Earth}\label{sec:sunandearth}

In this Section we examine the prospects of observing LOLIPs which are produced from DM annihilation.  The most promising sources to look for are regions where the DM density is enhanced, such as the Sun, Earth, sub-halos and Galactic Center (GC).   Out of those, only the Sun and Earth are suitable for detections in neutrino telescopes since the probability of LOLIPs to decay inside the detector is too small for more distant sources. Below we consider the constraints from SuperK, Fermi and  Milagro on LOLIPs originating from the Sun and Earth,  and study the reach of IceCube for discovering these particles.  We begin by shortly reviewing the formalism and relevant constraints before presenting our results.

\subsection{Production Rate}
\label{sec:ProdRate}

The production rate of LOLIPs in the Sun and Earth depends on the
 capture rate, which can vary by many orders of magnitude depending on
 the nature of the DM.  This topic has been studied extensively in the
 literature~\cite{Gould:1987ir,Jungman:1995df} and here we shortly review the necessary
 ingredients.

 To determine the rate one solves the evolution equation for the DM
 number density,
\begin{eqnarray}
  \label{eq:3}
  \dot N = C - C_A N^2\,.
\end{eqnarray}
Here $C$ is the capture rate and $C_A=\langle \sigma
v\rangle/V_{eff}$ is the thermally averaged annihilation cross
section, per effective volume, with $V_{eff} = 1.8\times
10^{26} (\mdm/\rm{TeV})^{-3/2}\unit{cm}^3$ for the
Sun~\cite{Griest:1986yu,Jungman:1995df} and $V_{eff} = 5.7\times 10^{22}(\mdm/\rm{TeV})^{-3/2}\unit{cm}^3$
for the Earth~\cite{Gould:1987ir,Jungman:1995df}.  Solving the above equation, the DM
annihilation rate in the Sun is found to be,
\begin{eqnarray}
  \label{eq:4}
  \Gamma = \frac{1}{2}C_AN^2 = \frac{1}{2}C\tanh^2(\sqrt{C C_A}t)\,,
\end{eqnarray}
where $t = 4.5\times 10^9$ years is the age of the solar system.
For sufficiently large capture rate and DM annihilation rate, the DM
number density saturates and the annihilation rate can be well
approximated by $\Gamma = C/2$.  This approximation holds for
$C\gtrsim\Csun^c = 3\times 10^{17}  (\mdm/\rm{TeV})^{-3/2}\unit{sec}^{-1}$
for the Sun and $C\gtrsim \Cearth^c =  10^{14}  (\mdm/\rm{TeV})^{-3/2}\unit{sec}^{-1}$
for the Earth, where we assumed the usual WIMP annihilation
cross-section, $\langle\sigma v\rangle = 3\times 10^{-26}
\unit{cm}^3\unit{sec}^{-1}$.  This approximation turns out to be rather good in the Sun
but not in the Earth where equilibrium is typically not reached.

The above solution depends both on the DM capture and annihilation
rate.  The former is model dependent and in particular depends on
whether the DM interacts elastically with the nucleus as assumed in many direct detection searches.  If there are more complicated interactions with nucleons, as in Inelastic DM~\cite{idm} or in other recent models~\cite{directdetectionavoidance}, the capture rate can be significantly modified.  In order to demonstrate the effects of a modified capture rate, we consider elastic scattering and inelastic scattering.
For the elastic case see~\cite{Gould:1987ir,Jungman:1995df} while the
inelastic case has been recently studied
in~\cite{Nussinov:2009ft,Menon:2009qj}.  In the inelastic scenario,
the DM states are split with a mass difference $\delta$, and
scattering off nuclei excites the light state into the heavier one.
While the net effect of inelastic scattering is to lower the capture
rate, its virtue is in relaxing the constraints on the DM-nucleon
cross-section arising from direct detection, by as much as three orders of
magnitude.  This allows for a significant enhancement in the
annihilation rate from the Sun. On the other hand the rate from the Earth shuts
down already for very low mass splittings, since elastic
scattering always dominates there.

We have repeated the analysis of~\cite{Nussinov:2009ft,Menon:2009qj}
and have reproduced both the elastic and inelastic capture rates.   
As a reference, it is convenient to express the capture rate as a
function of both the DM mass and splittings in a closed form.  We
find, 
\begin{eqnarray}
  \label{eq:14}
  C_i \simeq  \kappa_i\, f_i(\delta, \mdm) \left(\frac{1
      \unit{TeV}}{\mdm}\right)^{3/2} \left(\frac{\rho_{DM}t}{0.3\unit{GeV
        cm}^{-3}}\right)\left(\frac{250\unit{km
        s}^{-1}}{v_{\rm disp}}\right)^{3}
\left(\frac{\sigma^p_{\rm SI}}{10^{-43}\unit{cm}^2}\right)\,,
\end{eqnarray}
with $i={\odot,\oplus}$ refers to the Sun and the Earth cases. $\rho_{DM}$ is the DM density in the solar system, $v_{\rm
  disp}$ is the DM velocity dispersion and $\sigSI^p$ is the
spin-independent DM-proton cross-section which we assume (as is
typically the case) to dominate over the spin-dependent cross-section.
The constants $\kappa_i$ have values
\begin{eqnarray}
\label{eq:14b}
\hspace{2 cm}\kappa_{\odot} \simeq 2.2
\times   10^{20}\mathrm{s}^{-1}
\hspace{2 cm}
  \kappa_{\oplus} \simeq 6.6
\times   10^{10}\mathrm{s}^{-1}.\hspace{3 cm}
  \label{eq:15}
\end{eqnarray}
The functions $f_i$ are approximate functions fitted to
the inelastic results, and are given by,
\begin{eqnarray}
  f_\odot(\delta, \mdm) &=&  \frac{\tanh(1-\delta/(120\,\mathrm{keV})) +
    1}{\tanh(1) + 1}\,,
 \\
 f_\oplus(\delta, \mdm) &=& \delta(\delta -0\,\unit{keV})
\end{eqnarray}
The above fit is correct at the $20\%$ level for $\mdm \gtrsim
200\unit{GeV}$ and to about $50\%$ at $\mdm = 100$ GeV.  We note that
inherent uncertainties arising from the astrophysical uncertainties 
are significantly larger, thereby rendering the above approximation sufficient.

\subsection{Existing Constraints}
\label{sec:existing-constraints}

While a great deal of the parameter space for neutral LOLIPs is unexplored, some constraints already exists. ÊCollider constraints from low energy experiments have been covered in many different ways in~\cite{ArkaniHamed:2008qp, Han:2007ae,Essig:2009nc,reece,Aubert:2009pw,Bjorken:2009mm,pospelovFT}. ÊHowever, if DM particles can annihilate into LOLIPs there are additional indirect constraints on some part of the parameter space. ÊThese constraints roughly fall into three categories: existing neutrino telescope bounds, photon and electron fluxes from the Sun, and DM direct detection experiments. Ê
If the LOLIPs subsequently decay into muons, then SuperK bounds from up-going muon searches from both the Sun and the Earth are relevant~\cite{upgoingmuons,superkenergy}. ÊAs explained in Section~\ref{sec:experiments}, these searches are just a rough bound on the LOLIP scenario because utilize on neutrino induced muon events in SuperK. Ê

As long as the LOLIP effective lifetime is long enough such that they escape the Sun, it will produce decay products between the Sun and the Earth. These decays may give rise to signals which can be explored using experiments other than neutrino telescopes. ÊFor instance if the LOLIP decays to charged particles these particles will radiate photons on their way from the Sun to the Earth that can be detected by gamma ray searches such as Fermi or Milagro. ÊIf the LOLIPs decay products ultimately produce $e^{\pm}$, additional bounds coming from the Fermi measurement of the total $(e^++e^-)$ flux apply. Ê
Specifically for the Fermi $\gamma$ measurements we consider the differential $\gamma$ ray spectrum for the Sun, as was recently presented by the Fermi Collaboration~\cite{fermisun}, extending up to $10\,{\rm GeV}$ and fitted by a power law with a spectral index of -2.25. We require that the LOLIP contribution should not exceed twice the size of the error bars in any given bin. The same procedure is applied for the case of $(e^++e^-)$ Fermi data. On the other hand, the Milagro $\gamma$ data extends to much higher energies than those of Fermi. In order to compute the bounds we use the publicly available effective area and the exposure time to extract the total number of events expected in Milagro.  We then extract the 90\% CL intervals according to the procedure described in~\cite{Atkins:2004qr}. 
Before passing to the Direct Detection experiments, we further note that even though these bounds are directly applicable to the Sun, they also indirectly bound the production of LOLIPs in the Earth because the capture and annihilation rates in the two astrophysical objects are related.

The last type of bound from DM production of LOLIPs is from DM direct detection experiments. ÊThis bound comes from the fact that  any capture of DM in the Sun or Earth than requres a non-vanishing DM-nucleon interaction. ÊUsing the relation between the annihilation and capture rates we can then translate a point in the parameter space of lifetime and annihilation rate into a specific value for $\sigma_{\chi n}$. ÊRequiring that this interaction cross-section does not violate the constraints from the XENON10~\cite{xenon10} and CDMS~\cite{cdms} experiments, one can bound the parameter space. ÊIt should be noted that this translation is model dependent, e.g. it depends whether the scattering is elastic or inelastic.  Recently there have been many other models proposed that can also change the nature of DM direction detection~\cite{directdetectionavoidance}: We do not analyze the bounds for all these cases, but warn the reader of the many caveats in the existing direct detection constraints.

\subsection{Results}
\label{sec:results}

\begin{figure}[t] 
 \centering   \includegraphics[width=6in]{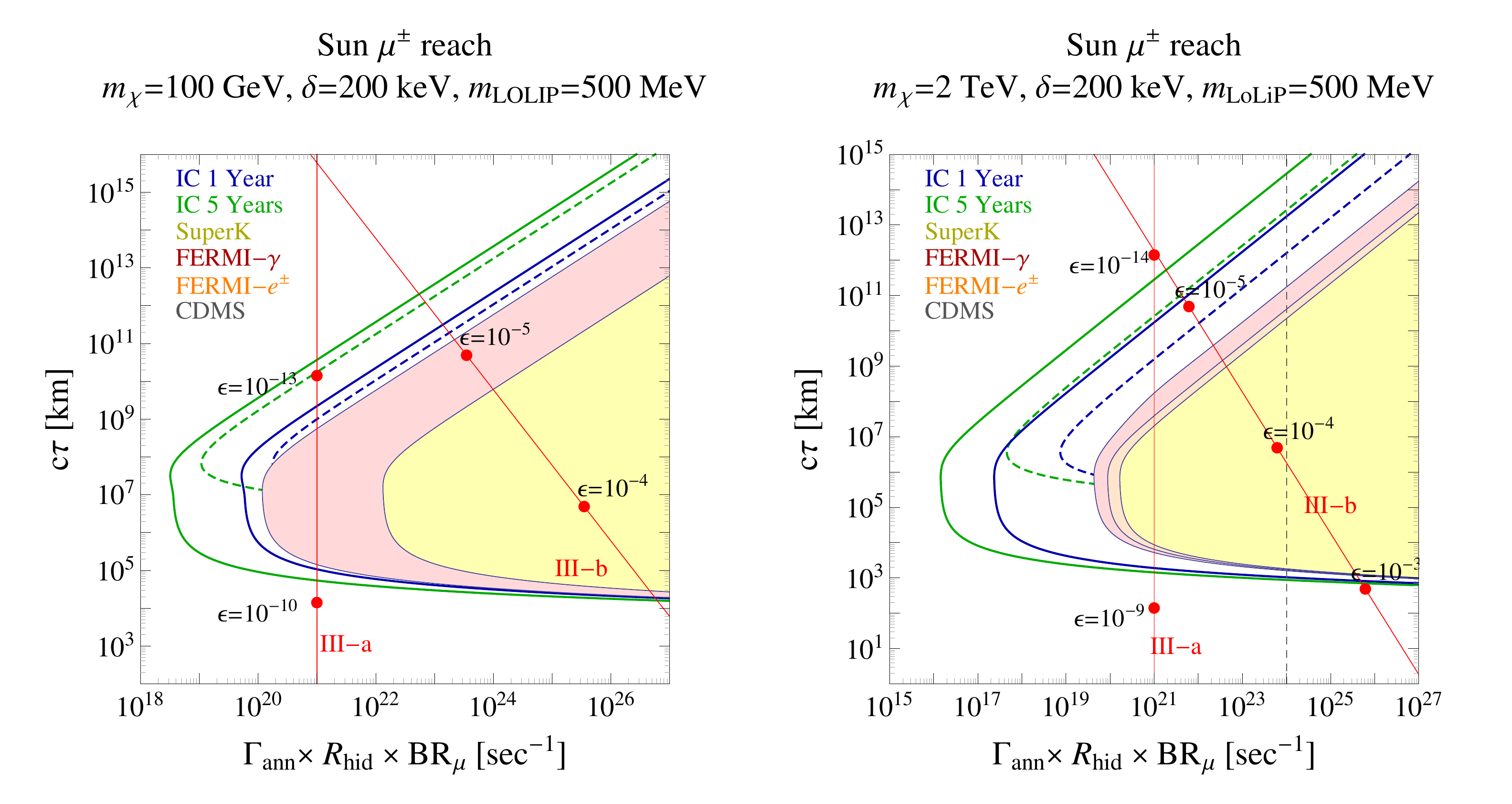}   \caption{The IceCube reach for detecting LOLIPs produced in the Sun and decaying into muons, as a function of the annihilation rate and decay length. On the x-axis, $R_{\rm hid} = \sigma({\rm DM\, DM}\rightarrow{\rm LOLIPs}) / \sigma_{\rm total}$ and $BR_\mu$ is the branching fraction of LOLIPs into muons.  $100\unit{GeV}$ and $2\unit{TeV}$ DM masses are shown, assuming DM interacts inelastically with mass splitting of $200\unit{keV}$ and decays into $500\unit{MeV}$ LOLIPs.  The shaded regions are excluded by SuperK (yellow), Fermi photon measurements (red) and  Fermi $(e^++e^-)$ measurement (orange).    The blue and green solid lines show the one and five year sensitivity of IceCube to measuring LOLIPs.  To demonstrate the sensitivity to the special di-muon events from LOLIP decays inside the detector, we show in dashed lines the one and five year discovery reach for such events only.   For reference two theory lines corresponding to models (III-a) and (III-b) described in Section~\ref{sec:models}  are shown.  Finally the dashed gray line is the current CDMS bound on the annihilation rate.  This constraint is irrelevant in the $100\unit{GeV}$ case.}
 \label{fig:sunMu}
\end{figure}

\subsubsection{Sun}

Let us now present the results for the discovery reach of LOLIPs from
the Sun.  In Fig.~\ref{fig:sunMu} we show the constraints which arise
from SuperK, Fermi and CDMS together with the detection reach in
IceCube for one and five years, in  the case of a LOLIP decaying into
muons.   These plots and the ones to follow show the sensitivity of IceCube in the decay length - annihilation rate plane.  The measured rate is suppressed with respect to the annihilation rate, by $R_{\rm hid} = \sigma({\rm DM\, DM}\rightarrow{\rm LOLIPs}) / \sigma_{\rm total}$ and by the specific branching fraction of LOLIPs studied.
Before proceeding, a few words on our treatment of the backgrounds are in order.

The IceCube curves represent the 95\% CL limits. We use the atmospheric neutrino flux~\cite{ph0611418} to estimate the background. For the case of LOLIPs decaying in proximity of the detector, we considered the expected atmospheric $\nu_{\mu}$ flux above $500\,{\rm GeV}$, independently of the LOLIP energy, because to the different Cherenkov light yield of a di-muon pair. On the other hand, for the events initiated by neutrinos from earlier muon decays, we consider the full atmospheric muon background above the detection threshold. We study the flux coming from a cone of $O(3^\circ)$ around the source of interest.
In the case of SuperK, we directly use the bounds provided for standard DM $\nu_{\mu}$ searches~\cite{upgoingmuons} and also consider the showering-muon analysis~\cite{superkenergy} to directly bound LOLIPs decays (including also the showering-$\mu$ fraction from $\nu_{\mu}$ events as outlined in~\cite{superkenergy,Hisano:2009fb}).

\begin{figure}[t] 
  \centering   \includegraphics[width=6in]{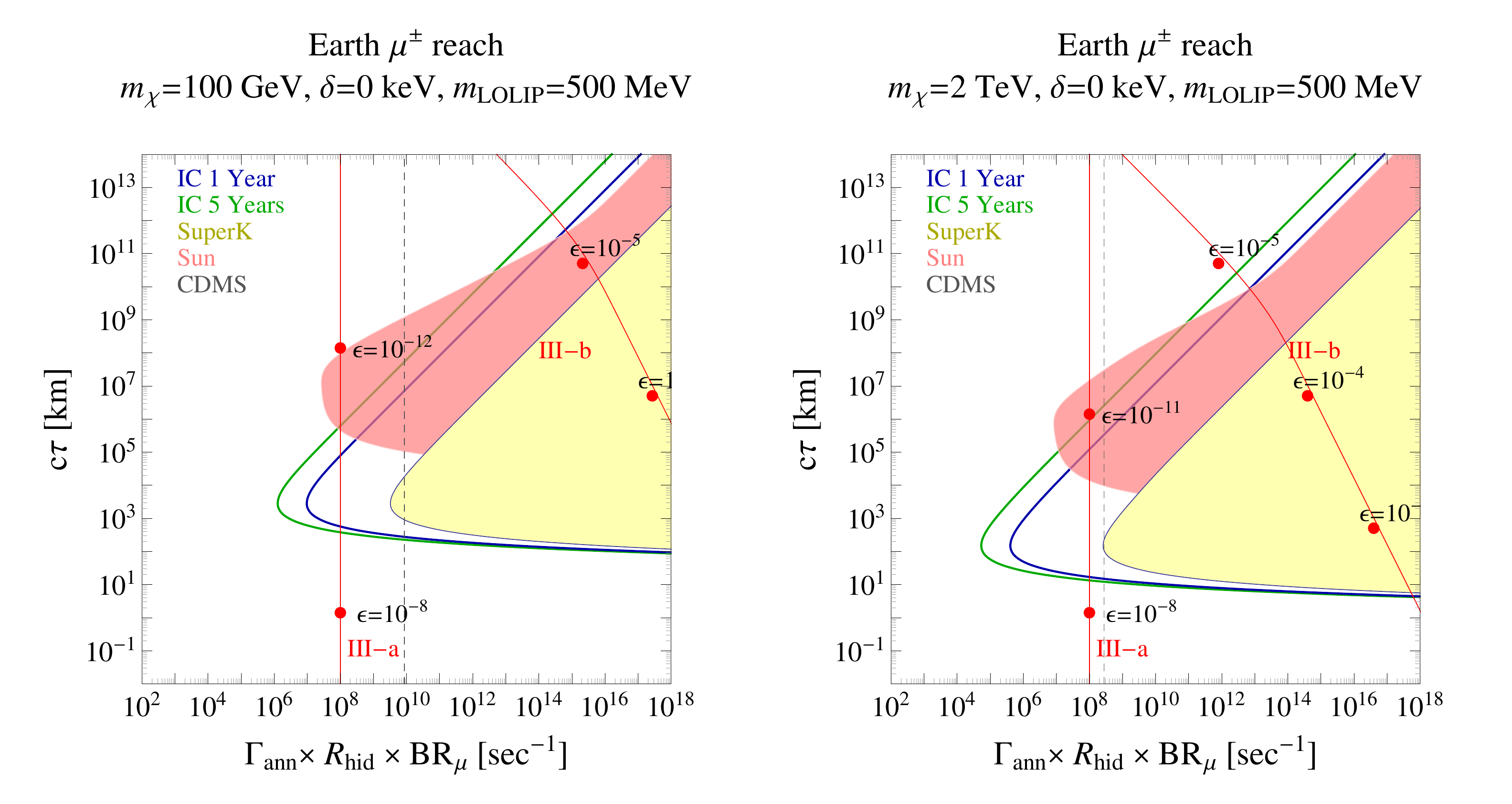} \caption{Similar to Figure~\ref{fig:sunMu} for the case of annihilation in the Earth.  The inelastic splitting is taken to be zero since the capture rate is highly suppressed otherwise. The red shaded region shows the constraints from the Sun, assuming that a non-vanishing annihilation rate at the Earth implies a related annihilation rate in the Sun.}
  \label{fig:EarthMu}
\end{figure}

In both plots of Figure~\ref{fig:sunMu} we have taken the LOLIP mass $\mlolip =
500$ MeV.  The main features of the figures are easy
to understand: The detection rate is maximized when the LOLIP decays
at or just in front of the detector. Given the LOLIP boost of order $\gamma \sim
\mdm/\mlolip$, one finds a linear decrease in the reach for lifetimes
above the critical value, $c\tau^c
\simeq m_{LOLIP}\cdot 1{\rm AU} / \mdm=0.5 {\rm GeV\, AU}/\mdm$, where AU$=1.5 \times 10^8$ km.   This linear
decrease is in accordance to Eq.~(\ref{surviveanddecay}).  Below $c\tau^c$,  we find that the sensitivity is roughly constant as long as the LOLIP escapes the Sun.  This can be understood as follows.  LOLIPs
that decay to muons before reaching the Earth produce neutrinos.  The
flux of these neutrinos is dominated by the decays occurring very close
to the Sun's surface and the probability of detecting them is
suppressed by the probability of the neutrinos to interact with the
detector,
\begin{eqnarray}
\label{eq:2}
P_{\nu-\rm int} \simeq (R_{\mu}+d) \times \rho N_A \times \sigma_{\nu N} \,.
\end{eqnarray}
Here $d$ is the length of the detector, $R_\mu$ is the muon range, $\rho$ is the density of matter
in the detector, (ice or water), $N_A$ is the Avogadro number and
$\sigma_{\nu N}$ is the Neutrino-nucleon cross-section.  Rather amusingly,
$P_{\nu-\rm int} \sim P_{\rm decay}$ for $\mdm \simeq 100$ GeV, which explains the flat
region in the plot as $c\tau$ drops below $c\tau^c$. The flat region stops at the value of $c\tau$ corresponding to the Sun radius.  The products
of  LOLIPs that do not escape the Sun are  not detected, as in this case the
muons stop inside the it giving rise to neutrinos of very low energies. 
In the plots, the dashed lines show the sole contribution of muons from LOLIPs decaying nearby the detector, while the solid lines include the
additional neutrinos produced before reaching the Earth.  As discussed in Sec.~\ref{sec:experiments}, the
tracks of di-muons produced inside the detector have
characteristic features which should allow one to differentiate them
from regular single muon events.
It is therefore conceivable that one can further reduce the backgrounds improving these results. Moreover, by measuring both the total flux and the di-muon fraction, a rather good estimation of $c\tau$ may be feasible.   
We postpone this analysis to future work.

Figure~\ref{fig:sunMu} also shows two calibrated ``theory'' lines, corresponding to models (III-a) and (III-b).  As discussed in
Section~\ref{sec:models}, Model (III-a) describes a hidden sector with a long
lived hidden gauge field which mixes with the SM through gauge
kinetic mixing.   We see that IceCube can probe extremely small
values of $\epsilon_V$.  Unless the DM interacts directly with the SM,
the capture rate and hence the annihilation rate from the Sun
vanishes for such low values.   Model (III-a) therefore assumes that the DM
is also coupled to the SM (see Section~\ref{sec:models}) and we assume
a DM-nucleon cross-section corresponding to an annihilation rate of
$10^{21}\unit{sec}^{-1}$, below the CDMS bound.
The same situation also applies to model (IV), hence this line can also be used to read off the numbers for this model as well, upon a trivial rescaling of $\epsilon_H=6\cdot 10^2 \epsilon_V $.
For the other theory line as in model (III-b), the lowest state
is a Higgs which can decay to SM fermions at one loop.  Here, the lifetime
is proportional to $\epsilon_V^{-4}$ and hence IceCube probes rather large
values of $\epsilon_V$. 

\begin{figure}[t] 
  \centering   \includegraphics[width=6in]{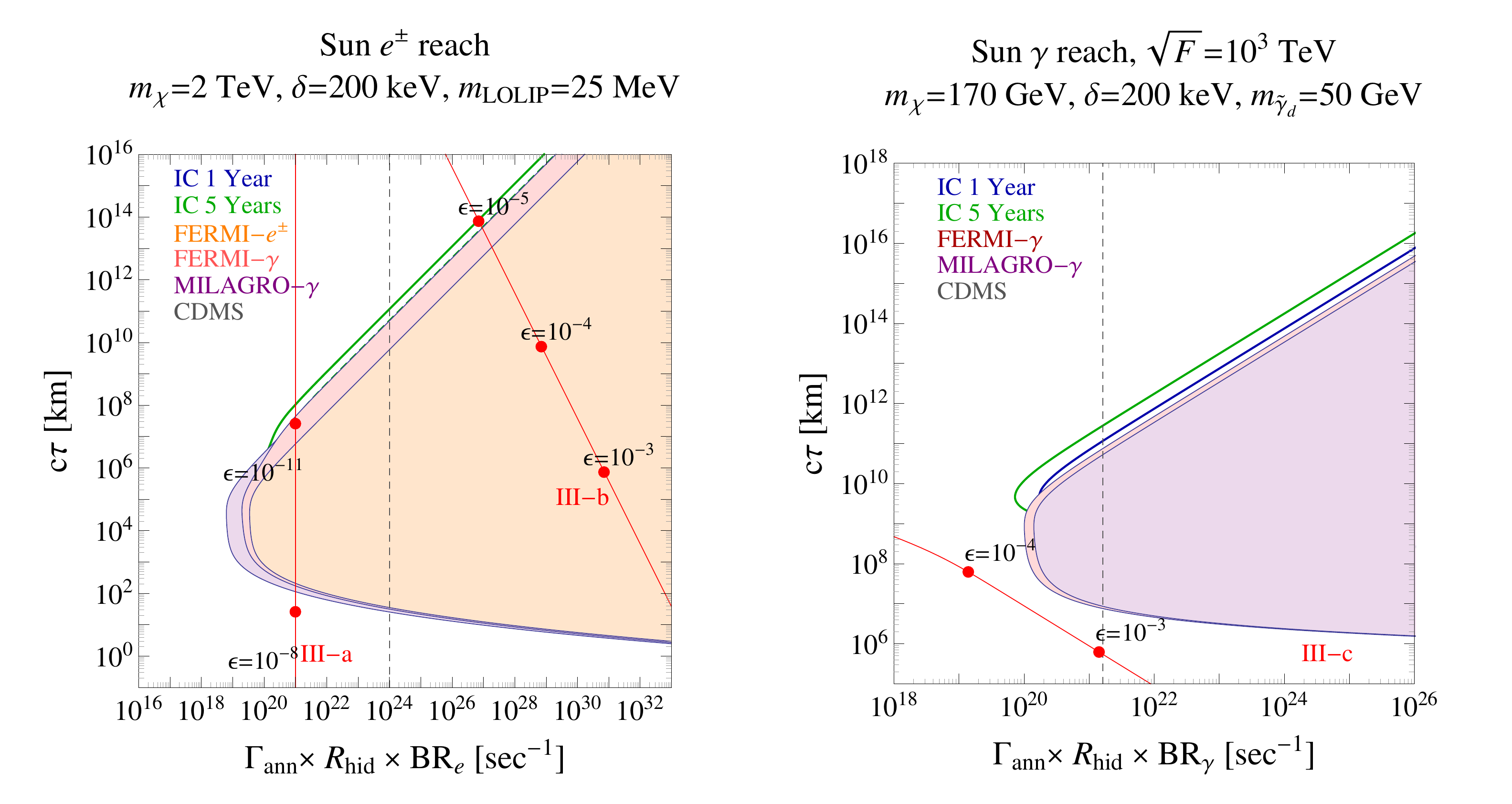} \caption{{\it Left:} The reach for inelastic DM annihilating in the Sun into LOLIPs with a mass of 25 MeV, which subsequently decays into $e^+e^-$.  For reference two theory lines corresponding to models (III-a) and (III-b) described in Section~\ref{sec:models}  are shown.  In addition to the constraints shown in Figure.~\ref{fig:sunMu}, we plot the constraint on the photon flux from Milagro which is insignificant in the muon case.   {\it Right:} The reach for annihilation of DM in the Sun into a LOLIP with a mass of 50 GeV, subsequently decaying via hard photon as in Model (III-c).  The axes in both planes are as in Figure~\ref{fig:sunMu}.}
  \label{fig:SunEG}
\end{figure}

In Figure.~\ref{fig:SunEG}a we plot the case for $\mdm=2$ TeV where
LOLIPs decay predominantly into electrons.   This would occur when
the LOLIP is very light, below twice the muon mass.  We therefore
take  $\mlolip=25\,$MeV.  As discussed in the previous Section, the
detection of electrons is different since they interact with
the detector promptly to produce a localized signal rather than a long track. Therefore the analysis should also be indicative of the reach for other final states producing electromagnetic or hadronic showers.
Unlike the muon case, we do not consider the SuperK bounds since the photons constraints from Fermi are comparatively much stronger than in the muon case. This is partly due to the fact that the
background from atmospheric electron neutrinos is
larger by roughly two orders of magnitude~\cite{IceCubePDD} than in the case of muons because of poorer directionality reconstruction for showers compared to $\mu$ tracks.  Unless a more detailed analysis is performed and better ways of background rejection are found, the bounds from Fermi gamma ray searches cover almost the full reach of IceCube and no discovery can be made in this case.
   In Fig.~\ref{fig:SunEG}b we also plot the IceCube reach for the case where LOLIPs decay into hard photons, and include a theory line for the model (III-c) described in Section~\ref{sec:models}. As one can see, Milagro gamma rays constraints cover most of the IceCube reach.

  \begin{figure}[t] 
  \centering   \includegraphics[width=6in]{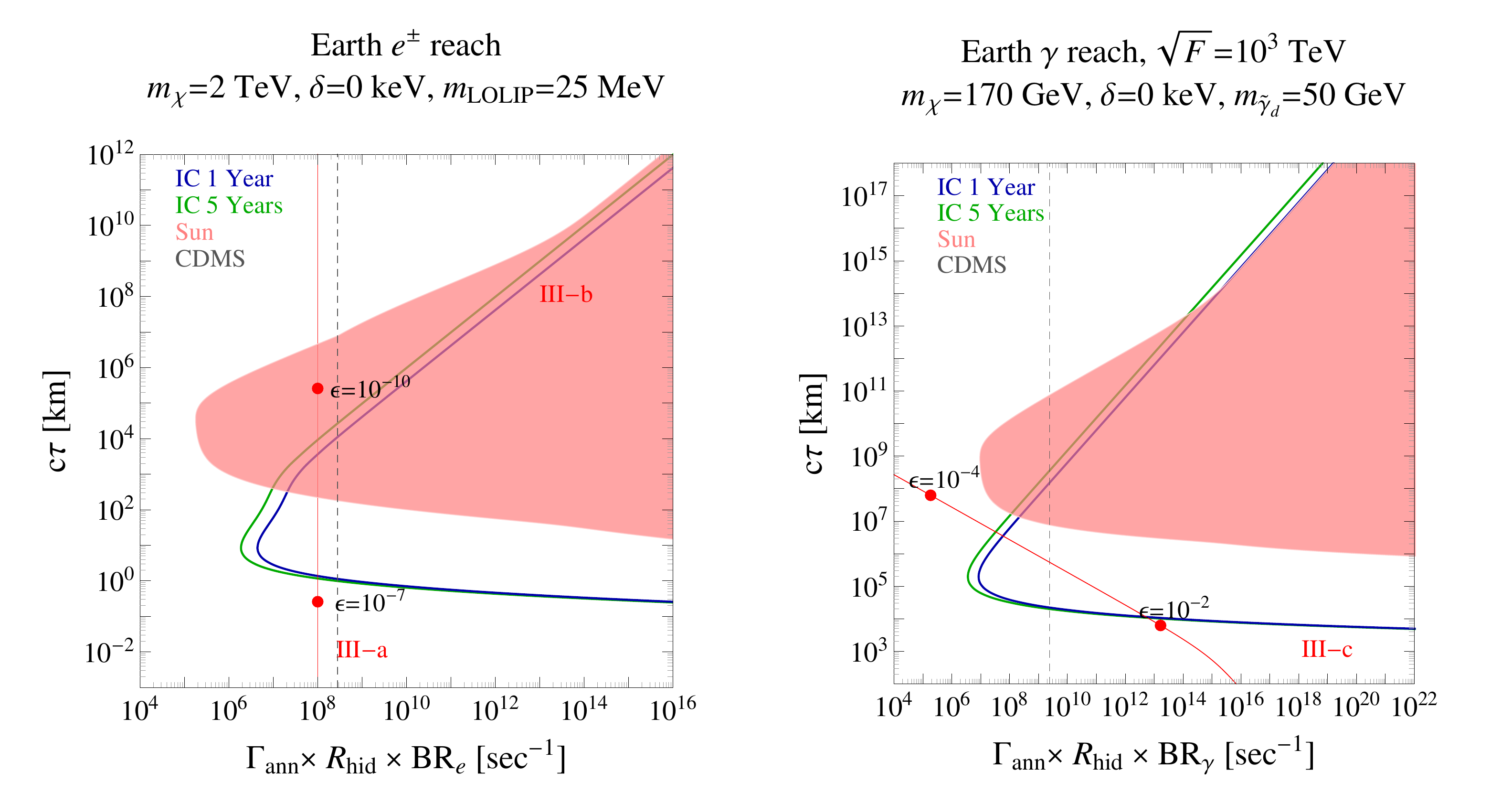} \caption{{\it Left:} As in the left panel of Figure~\ref{fig:SunEG}, only computed for annihilation in the Earth, no inelastic scattering, and DM mass of $2\unit{TeV}$.  {\it Right:} Similar to the right panel of Figure~\ref{fig:SunEG}.  The red region in both plots show the constraint from the Sun, assuming that a non-vanishing annihilation rate at the Earth implies a related annihilation rate in the Sun.}
  \label{fig:EarthEG}
\end{figure}

\subsubsection{Earth}

Let us now turn to the Earth.  Due to the different scales involved, the search for LOLIPs arising from DM annihilations at the center of the Earth enables one to probe  different lifetimes, of order $c\tau \sim R_\oplus / \mdm$.    As opposed to the Sun, LOLIPs that decay before reaching the detector stop instantly and therefore if any flux is measured, it must be directly related to LOLIPs decaying nearby IceCube.   Furthermore, as discussed earlier, the capture rate drops quickly as the mass splittings between DM states is increased.  Consequently, we concentrate on the elastic scattering region. 
In Figs.~\ref{fig:EarthMu} and~\ref{fig:EarthEG} we present similar plots to those of the Sun.    We take $\delta =0\,{\rm keV}$ and  plot the reach for IceCube as well as constraints from CDMS and SuperK (for the muon case). We also plot the strongest constraint from the Sun, which corresponds to the Fermi and Milagro bounds on the gamma ray flux. This assumes that the capture in the Earth would occur in conjunction with capture in the Sun.   As before, in the case of $e^\pm$ and $\gamma$ final states we do not include the SuperK bound for decays into electrons. This is because of the much larger backgrounds compared to muon searches, that weakens the bounds.  It is apparent that the allowed region that can be probed by IceCube is significantly smaller for the Earth case.  Nevertheless, such measurements allow us to cover a different region of parameter space.

\subsubsection{Comparison to Direct Detection Bounds}

\begin{figure}[t] 
 \centering
 \includegraphics[width=6in]{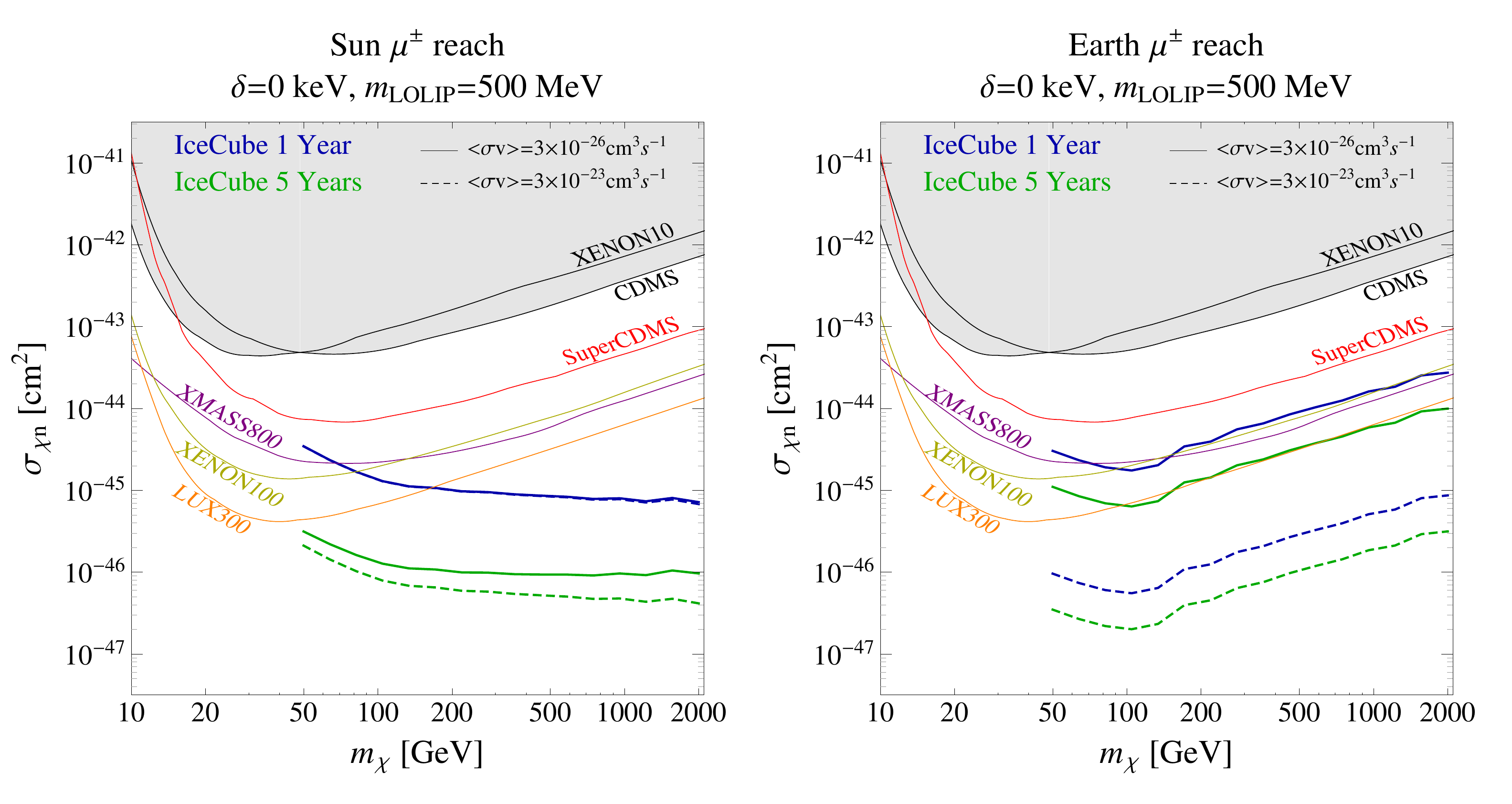}
\caption{In both the left and right panels of this figure we plot the reach for IceCube in the standard $m_\chi$ vs $\sigma_{\chi n}$ plane used for plotting constraints from direct detection experiments.  The procedure for translating from $\Gamma$ and $c\tau$ to $m_\chi$ and $\sigma_{\chi n}$ is described in the text.  Additionally we plot the current bounds from CDMS and XENON10 and the projected sensitivities for upcoming DM direct detection experiments. }
\label{fig:mSigma}
\end{figure}

It is interesting to understand whether the region in reach of IceCube can be ruled out by direct detection experiments in the near future.   The answer is presented in Fig.~\ref{fig:mSigma}, where we plot the maximal sensitivity of IceCube to the DM-nucleon cross-section as a function of the DM mass.   To extract this plot we pick, for a given DM mass, the optimal lifetime that would allow for the lowest annihilation rate.  We translate it to the nuclear cross-section, assuming a particular DM annihilation cross-section, as outlined in Eqs.~\eqref{eq:14},\eqref{eq:14b}.  The curves are drawn for the $\mu^\pm$ final states both for the case of the Sun and the Earth. For concreteness we fixed the LOLIP mass to $500\,{\rm MeV}$.   Since we scan over the LOLIP lifetime to achieve maximal sensitivity, the effect on Fig.~\ref{fig:mSigma} of changing $m_{\rm LOLIP}$ is only subleading.

The solid blue(green) line corresponds to the one(five) year reach for IceCube in the case of a WIMP annihilation cross-section, $\sigma=3\times 10^{-26}\unit{cm}^3\unit{s}^{-1}$ .   The dashed lines correspond to $\sigma=3\times 10^{-23}\unit{cm}^3\unit{s}^{-1}$  which is the cross-section required to explain the recent cosmic-ray anomalies~\cite{us}. Nucleon scattering cross-sections of $10^{-46} {\rm cm}^2$ can be achieved in 5 years in the case of the Sun irrespectively of the annihilation cross-section, while the Earth can reach that level only if DM annihilation is enhanced~\cite{Delaunay:2008pc}. We compare the results with the given and future constraints from direct detection. As one can see, for the case of the Sun, IceCube comparatively performs better at higher energies. We stress that the comparison is made for the elastic, $\delta=0\,\unit{keV}$, case.   For inelastic scattering, constraints from direct detection are significantly weaker and the corresponding IceCube reach (modestly weaker as well) can be easily inferred from the approximate formulas we have given.
Therefore, we conclude that IceCube will have a reach better or comparable to that of direct detection experiments and will provide an additional avenue to test non-standard DM scenario.

\begin{figure}[t] 
  \centering
  \includegraphics[width=6in]{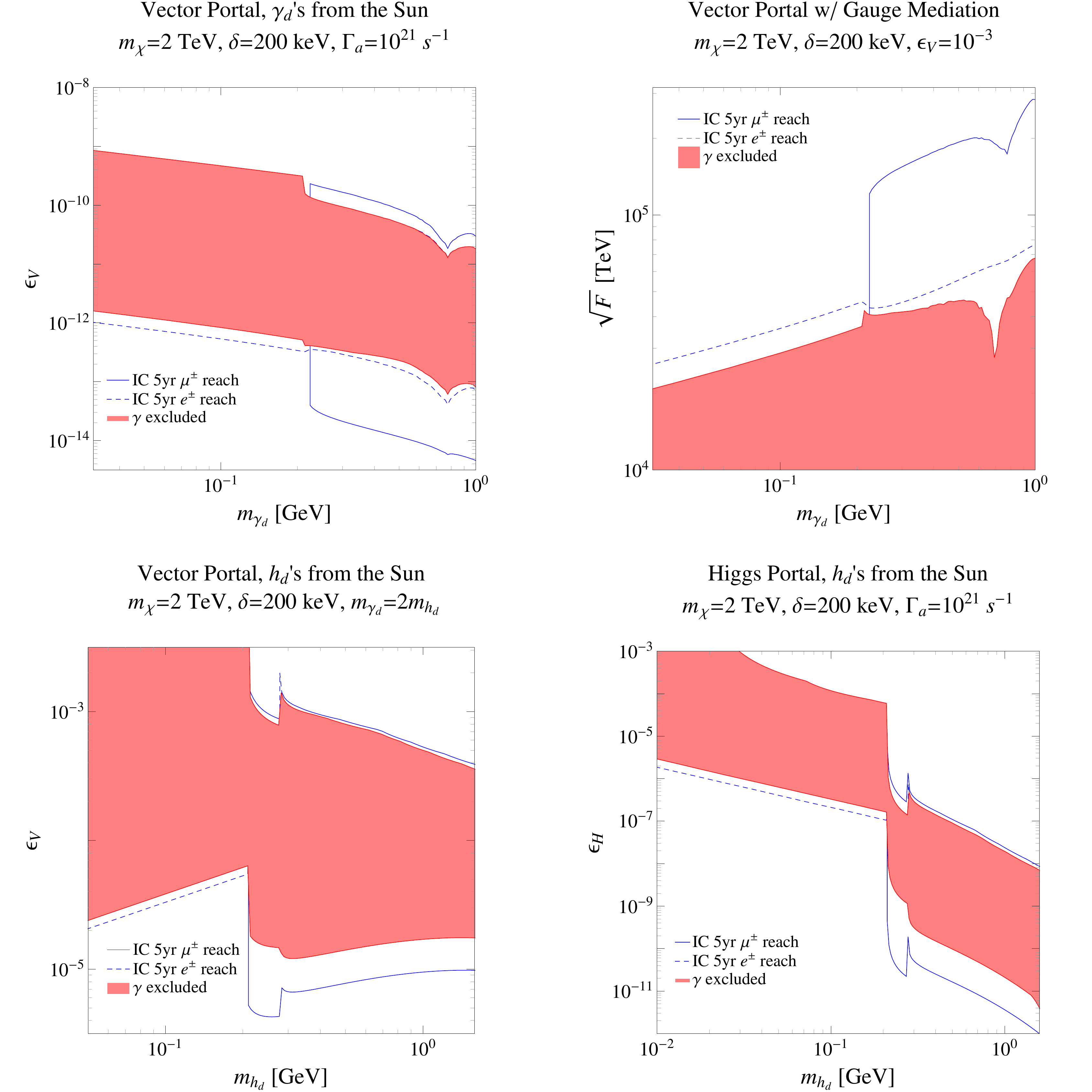}
 \caption{The current bounds and the IceCube reach in the parameter space of four benchmark models presented in Section~\ref{sec:models} are shown based on the results summarized in this section. Upper left: model (III-a). Upper right: model (III-d). Lower left: model (III-b). Lower right: model (IV).  The red band correspond to the photon exclusion region extracted from Fermi and Milagro data. }
 \label{fig:theory}
\end{figure}

\subsubsection{Constraints on Models}

Finally we conclude this Section by presenting the implications of the results described here for the models of Section~\ref{sec:models}.
This is illustrated in Figure~\ref{fig:theory} for the models (III-a,b,d) and (IV). The experimental reach and the current constraints in terms of the fundamental parameters of each model are shown. On the top-left panel we plot the Sun reach for a $2\unit{TeV}$ DM annihilating into a vector portal described in model (III-a), assuming mass splittings of $200\unit{keV}$ and $\epsilon_V$-independent annihilation rate of $10^{21}\,{\rm s}^{-1}$.   On the top-right panel we show a similar plot for the vector portal model, (III-d), where the LOLIP is the gaugino with a lifetime controlled by the SUSY-breaking scale.  The left-bottom panel is for the $2\unit{TeV}$ DM with the light higgs in the vector portal, and finally the bottom-right is the higgs portal, model IV.  
In all plots the excluded part of the parameter space due to the Fermi photon measurement is shown by the red shaded region.  

By changing the mass of the particle decaying back to the SM (which, except for model (III-d) corresponds to the LOLIP mass) the branching fractions into final state particles are changing. In particular in the case of models (III-a) and (III-d) the final states are selected by the requirement that the mediator couples to the electric charge, while in the other two cases by the requirement that the partial widths are proportional to $m_f^2$. For mediators lighter than 1 GeV this results in a combination of $e^+e^-$, $\mu^+\mu^-$ and $\pi^+\pi^-$, $\pi^0\pi^0$ final states. We estimate the admixtures by the measurement of the $e^+e^-\rightarrow {\rm hadrons}$ cross-section for the case of the electric charge, while in the other models we estimate the contribution of pions using chiral perturbation theory. All the kinks and drops shown in these plots can be readily understood in terms of the relative importance of these channels.
It is notable that we do not extend the plots for masses above $1\unit{GeV}$ (where low energy direct probes lose their sensitivity).  The reason for this is that while allowed regions  may still exist up to masses of the order of few GeV (photon constraints definitely becomes very stringent somewhere before 8~GeV~\cite{us}), QCD uncertainties do not allow for a satisfactory theoretical or experimental handles to determine the possible SM final states and their relative branching fractions. 

\section{Conclusions}\label{sec:conclusions}

In this paper we have studied the prospects of detecting neutral long lived particles (LOLIPs) with decay lengths ranging between $1\unit{km}$ and $10^{15}\unit{km}$. Ê ÊSuch long lived states may arise in several scenarios such as GMSB models and hidden sectors weakly coupled to the SM. Ê Previous studies focused on significantly shorter lifetimes, keeping many theoretical models out of the experimental reach. ÊWe stress that the added value of the approach discussed in this paper, is that it not only improves on the sensitivity to low scale weakly coupled sectors, but also it allows for detection of hidden sectors with mass scales significantly larger than $1\unit{GeV}$, such as the model (III-d) of Section~\ref{sec:models}.

To obtain sensitivity to such long lifetimes, we studied neutrino telescopes employing their large scale detectors. ÊFor the production mechanism, we considered the LHC and DM annihilations into LOLIPs in the Sun or the Earth. Ê ÊAttempting to directly probe LOLIPs produced at the LHC and detected at Antares or KM3NeT proved not feasible due to the low rates. ÊFuture improvements on the luminosity at the LHC and/or on the energy threshold at KM3NeT may render this possibility viable. ÊConversely, designing small scale detectors closer to the LHC may allow one detect LOLIPs which show up as MET at the LHC. ÊWe postpone the study of this possibility to future work~\cite{workinprogress}.
On the other hand, indirect detection of LOLIPs from the Earth and Sun seems promising, depending on the final states into which the LOLIPs decay. Ê In particular, if LOLIPs predominantly decays into di-muons, a significant Êregion of the parameter space previously unexplored can be probed by IceCube. ÊConversely, the prospects of detecting final states which shower inside the detector, such as electrons or photons, strongly depend on the capabilities of reducing the atmospheric neutrino background. Ê

Finally, in relation to DM annihilating into LOLIPs, Êwe have compared the capabilities of the neutrino telescope techniques, to those of direct detection. We found that for models which accommodate long lived states, neutrino telescopes can detect annihilation fluxes from the Sun corresponding Êto nucleon cross-sections beyond the reach of upcoming direct detection experiments.

Note Added:  While this work was in writing, related work appeared~\cite{Batell:2009zp}.

{\bf Acknowledgments.}
We thank N.~Arkani-Hamed, J.~Ruderman, P.~Schuster, M.~Strassler, N.~Toro, J.~Uscinski and I.~Yavin for useful conversations.   PM, MP, and TV would like to thank the Galileo Galilei Institute for Theoretical Physics in Florence, Italy where part of this work was completed.  MP would like to thank the Aspen Center for Physics where part of this work was completed. PM and TV are supported in part by DOE grant DE- FG02-90ER40542. MP is supported in part by NSF grant PH0503584.

\end{document}